\documentclass[preprintnumbers, prd, onecolumn, compress,showpacs, floatfix,
preprintnumbers, a4paper, superscriptaddress,nofootinbib]{revtex4}
\usepackage{amsmath}
\usepackage {amssymb}

\def\tref{h_{(\rm r)}}
\def\Lag{{\mathcal L}{}}
\def\ombol{{\stackrel{\circ}{\omega}}{}}

\begin{document}

\title[The covariant formulation of $f(T)$ gravity]{The covariant formulation of $f(T)$ gravity}

\author{Martin Kr\v{s}\v{s}\'ak}
\email{krssak@ift.unesp.br}
\address{Instituto de F\'{\i}sica Te\'orica,
Universidade Estadual Paulista \\
Rua Dr.\ Bento Teobaldo Ferraz 271, 01140-070 S\~ao Paulo, SP, Brazil}

\author{Emmanuel N. Saridakis}
\email{Emmanuel\_Saridakis@baylor.edu}
\address{CASPER, Physics Department, Baylor University, Waco, TX 76798-7310, USA} 
\address{Instituto de F\'{\i}sica, Pontificia
Universidad de Cat\'olica de Valpara\'{\i}so, Casilla 4950,
Valpara\'{\i}so, Chile}

\begin{abstract}
We show that the well-known problem of frame dependence and violation of local 
Lorentz invariance in the usual formulation of $f(T)$ gravity is a consequence of 
neglecting the role of spin connection. We re-formulate $f(T)$ gravity 
starting, instead of the ``pure-tetrad''  teleparallel gravity, from the covariant 
teleparallel gravity, using both the tetrad and the spin connection as dynamical 
variables, resulting in the fully covariant, consistent, and frame-independent, version 
of $f(T)$ gravity, which does not suffer from the notorious problems of the usual, 
pure-tetrad, $f(T)$ theory. We present the method to extract solutions for the most 
physically important cases, such as  the Minkowski, the FRW and the 
spherically-symmetric ones. We show that in the covariant $f(T)$ gravity we are allowed 
to use an arbitrary tetrad in an arbitrary coordinate system along with the corresponding 
spin connection, resulting always to the same physically relevant field equations.

\end{abstract}


\maketitle

\section{Introduction} 

Modified gravity \cite{Nojiri:2006ri,Capozziello:2011et} is one of the two main 
approaches that one can follow in order to describe the two accelerated phases of 
expansion, at early and late times respectively (the other one is to introduce the 
concept of dark energy in the framework of general relativity 
\cite{Copeland:2006wr,Cai:2009zp}), 
which moreover has the additional motivation of alleviating the difficulties towards the 
quantization of gravity and improving its UltraViolet behavior 
\cite{Stelle:1976gc,Biswas:2011ar}. 
However, even if one decides to take the serious step of modifying gravity, there is 
still the question of what formulation of gravity to modify. Most of the works in the 
literature start from the usual, curvature-based formulation, and modify/extend 
the Einstein-Hilbert action, with the simplest example being the $f(R)$ paradigm 
in which the Lagrangian is considered to be a non-linear function of the 
curvature scalar \cite{DeFelice:2010aj,Nojiri:2010wj}.

However, one could start from the Teleparallel Equivalent of General Relativity 
(TEGR) 
\cite{ein28,Unzicker:2005in,Sauer,Moller,Hayashi:1967se,Cho:1975dh,Hayashi:1977jd,
Hayashi79,Maluf:1994ji, Arcos:2005ec,Pereira.book}, in which gravity is described through 
torsion and the gravitational Lagrangian is the torsion scalar $T$, and construct various 
extensions. In these lines, the simplest torsional modification is the  $f(T)$ paradigm, 
in which the Lagrangian is taken to be a non-linear function of $T$ 
\cite{Ferraro:2006jd,Bengochea:2008gz,Ferraro:2008ey,Linder:2010py}. The crucial issue is 
that although TEGR 
coincides completely with general relativity at the level of equations, $f(T)$ is 
different from $f(R)$ gravity, with novel features (amongst others note the significant 
advantage that the field equations of $f(T)$ gravity are of second order while those of 
$f(R)$ are of fourth order) and interesting cosmological 
implications, and that is why it has gained a lot of interest in the literature 
\cite{Chen:2010va,Wu:2010mn,Bengochea001,Dent:2011zz,Zheng:2010am,Bamba:2010wb,Cai:2011tc,
Sharif001,Li:2011rn,Capozziello:2011hj,Daouda:2011rt,Wu:2011kh,Wei:2011aa,
Atazadeh:2011aa,Farajollahi:2011af,Karami:2012fu,Cardone:2012xq,Jamil:2012ti,Ong:2013qja,
Amoros:2013nxa,Nesseris:2013jea,Bamba:2013ooa,Kofinas:2014owa,Harko:2014sja,Haro:2014wha,
Geng:2014nfa, Hanafy:2014ica,Darabi:2014dla,Capozziello:2015rda,Nashed:2015pda}.

Unfortunately, in the standard formulation of $f(T)$ gravity local Lorentz invariance is 
either completely absent or strongly restricted \cite{Li:2010cg,Ferraro:2014owa}, due to 
the strong imposition made in 
\cite{Ferraro:2006jd,Bengochea:2008gz,Ferraro:2008ey,Linder:2010py} that the spin 
connection vanishes. Although this assumption has a good motivation, namely to make the 
theory simpler in order to be able to extract solutions, and although it is based on the 
fact that the spin connection is not a tensor and under a local Lorentz transformation it 
transforms non-covariantly and thus it is always possible to transform to a frame that 
it is zero, in general this assumption makes the theory frame-dependent 
since a solution of the field equations depends on the choice of the frame 
\cite{Li:2010cg}. 
One can neglect this issue and investigate solutions in particular 
frames (this is in analogy with the investigation of electromagnetism in the particular 
class of inertial frames), however strictly speaking the problem is there and will 
become obvious when questions about frame transformations and Lorentz invariance are 
raised, which is usual for instance in the case of spherically symmetric solutions. 

The feature of local Lorentz invariance violation is clearly a deficit of the standard 
formulation of $f(T)$ gravity. While this problem could be avoided including higher 
derivative terms \cite{Bahamonde:2015zma,Junior:2015kia}, the resulting theories loose 
the 
most attractive feature of $f(T)$ theories, which is that the field equations are of 
second order. In the present work we desire to re-formulate $f(T)$ gravity in order to be 
fully covariant, and this will be achieved by relaxing the strong assumption of setting 
the spin connection to zero. Hence, we obtain a theory that has both attractive features: 
preserves local Lorentz symmetry and the field equations are  of 
second order.

The outline of this paper  is as follows. In Section~\ref{tegr} we briefly introduce the 
covariant formulation of teleparallel gravity, keeping both the tetrad and the spin 
connection as dynamical variables, and focusing on the role of the inertial 
effects in the theory, explaining why the field equations are not affected by them. Based 
on that, in Section~\ref{ftsec} we construct the covariant and consistent version of  
$f(T)$ theories, deriving the field equations which include the spin connection. In 
Section \ref{solutions} we present the method of extracting solutions in such theories,  
illustrating it in the most physically relevant cases such as the Minkowski, the 
Friedmann-Robertson-Walker (FRW) and the spherically-symmetric ones. Lastly, in Section 
\ref{conclusions} we summarize our results.

\section{Covariant teleparallel gravities}
\label{tegr}

In this section we discuss on the Lorentz invariance and the various versions of 
teleparallel gravity. In a first subsection we present the covariant formulation of 
teleparallel gravity, while in a second subsection we present the covariant formulation 
of the teleparallel equivalent of general relativity, in which one keeps both the tetrad 
and the spin connection as dynamical variables.

\subsection{General covariance, tetrad and spin connection}

Let us first start the discussion by the ordinary teleparallel gravity and its origins. 
In his first relevant papers, Einstein was motivated by the observation that a tetrad 
has 16 independent components, of which only 10 are needed to determine the metric tensor 
and hence describe gravity, and thus the additional 6 degrees of freedom could 
describe the electromagnetic field \cite{ein28,Unzicker:2005in,Sauer}.
However, later on  \cite{Sauer,Moller,Hayashi:1967se,Cho:1975dh,Hayashi:1977jd,
Hayashi79} it was realized that the six 
additional degrees of freedom in the tetrad were actually related to the possible ways of 
choosing the observer, and therefore related to the inertial effects instead of 
electromagnetism, and the corresponding theory was named teleparallel 
gravity (due to the way one ``parallelizes'' the tetrads from a ``distance''). We mention 
that in this formulation of teleparallel gravity one still uses the original Einstein's 
definition of teleparallelism, which in our modern language can be defined as a geometry 
where the spin connection vanishes identically. Hence, the only dynamical variable is the 
tetrad, and therefore we refer to it as \textit{pure tetrad teleparallel gravity}. The 
disadvantage with such a construction is that the torsion tensor is ``effectively 
replaced'' by the coefficients of anholonomy, that are not tensors under local Lorentz 
transformations, which is the cause of violation of local Lorentz symmetry in this 
theory. 
However, it turns out that these violations do not affect the field equations, which are 
still invariant under local Lorentz transformations \cite{Obukhov:2006sk}, which ensures 
that the metric tensor obtained from pure tetrad teleparallel gravity is the correct one 
(see \cite{Maluf:2013gaa} for a recent review on this subject).
 
Later on it was realized that if the teleparallel geometry is defined more generally as 
a geometry with zero curvature, the most general spin connection that satisfies this 
requirement is the \textit{purely inertial spin connection}, analogous to a pure gauge 
connection in gauge theories. Such a connection vanishes only in a very special class of 
frames called ``proper frames'', which are the frames in which inertial effects are 
absent. 
This is a very subtle, but important difference with the pure tetrad formulation, where 
the spin connection is considered to vanish in \textit{all} frames. The main advantage of 
this approach is that if the purely inertial connection is used, teleparallel gravity has 
manifest local Lorentz invariance. We refer to this theory as the \textit{covariant 
teleparallel gravity}. This approach was pioneered in \cite{Obukhov:2002tm} in the 
framework of metric-affine theories \cite{Hehl:1994ue}, and then investigated further 
in \cite{Obukhov:2006sk,Lucas:2009nq} and fully adopted in \cite{Pereira.book}. 

The resulting, covariant, teleparallel gravity seems to be more complicated than the pure 
tetrad one, due to the appearance of the spin connection, however, as it turns out, one 
can find a self-consistent method to solve the field equations and determine the spin 
connection \cite{Krssak:2015rqa}. In addition to having the theory that respects the 
crucial local Lorentz symmetry, we are able to separate the gravitational from inertial 
effects and remove the InfaRed divergences from the action 
\cite{Krssak:2015rqa,Krssak:2015lba}.

\subsection{Teleparallel Equivalent of General Relativity}

Having discussed above the general idea of covariantizing teleparallel gravity, in this 
subsection we present the covariant formulation of Teleparallel Equivalent of General 
Relativity (TEGR). In such a formulation the fundamental variables are the tetrad 
$h^a_{\ \mu}$ and the spin connection $\omega^a_{\ b\mu}$. The tetrad is a set of four 
orthonormal vectors that represents a frame of reference for the physical observer, and 
is 
related to the metric tensor through
\begin{equation}
g_{\mu\nu}=\eta_{ab}h^a_{\ \mu}h^b_{\ \nu}, \label{metdef}
\end{equation}
where $\eta_{ab}=\text{diag}(1,-1,-1,-1)$ (we use the  notation where the 
Latin letters denote the tangent space indices, while Greek letters denote the 
spacetime indices).

The spin connection determines the torsion and curvature tensors, which in return
completely characterize the spin connection, through
\begin{equation}
T^a_{\ \mu\nu}(h^a_{\ \mu},\omega^a_{\ b\mu})=
\partial_\mu h^a_{\ \nu} -\partial_\nu h^a_{\ \mu}+\omega^a_{\ b\mu}h^b_{\ \nu}
-\omega^a_{\ b\nu}h^b_{\ \mu},
\label{torr}
\end{equation}
and
\begin{equation}
R^{a}_{\,\,\, b\mu\nu}(\omega^a_{\ 
b\mu})= \partial_\mu\omega^{a}_{\,\,\,b\nu}-
\partial_\nu\omega^{a}_{\,\,\,b\mu}
+\omega^{a}_{\,\,\,c\mu}\omega^{c}_{\,\,\,b\nu}-\omega^{a}_{\,\,\,c\nu}
\omega^{c}_{\,\,\,b\mu}\,.
\label{curvv}
\end{equation}
In general relativity one uses the Levi-Civita connection $\ombol^a_{\ b\mu}$, which by 
construction gives vanishing torsion, and thus all the information of the gravitational 
field is embedded in the curvature (Riemann) tensor, while the gravitational Lagrangian 
is the curvature (Ricci) scalar. On the other hand, in TEGR one uses the teleparallel 
(Weitzenb\"{o}ck) spin connection,
which by construction gives vanishing curvature, and thus all the information of the 
gravitational field is embedded in the torsion tensor, while the gravitational 
Lagrangian is the torsion scalar. In particular, in TEGR the dynamics is derived from 
the Lagrangian density\footnote{For convenience, in this manuscript we follow the 
conventions  used in TEGR literature, that slightly differ from those in $f(T)$ 
gravity works by a factor of 2. Therefore, the superpotential in $f(T)$ gravity is 
usually defined as one half of our definition (\ref{superp}).}
\begin{equation}
\Lag=\frac{h}{4\kappa}T,
\label{tegraction}
\end{equation}
 with $h=\det h^a_{\ \mu}$ and $\kappa=8\pi G$  the gravitational constant, where
\begin{equation}
T= T^a_{\ \mu\nu}S_a^{\ \mu\nu}
\label{tscalar}
\end{equation}
is the torsion scalar constructed by contractions of the torsion tensor, with
\begin{equation}
S_a^{\ \mu\nu}=K^{\mu\nu}_{\ \ a}
-h_a^{\ \nu}\,T^{\alpha \mu}_{\ \ \ \alpha}
+h_a^{\ \mu}\,T^{\alpha \nu}_{\ \ \ \alpha}
\label{superp}
\end{equation}
the superpotential and  $K^{\mu\nu}_{\ \ a}$ the contortion tensor defined as 
\begin{equation}
K^{\mu\nu}_{\  \ a}=\frac{1}{2}
\left(
T^{\ \mu\nu}_{a}
+T^{\nu\mu}_{\ \ a}
-T^{\mu\nu}_{\  \ a}
\right).
\label{contortion}
\end{equation}
 The field equations are obtained by varying the gravitational Lagrangian 
(\ref{tegraction}) and 
the matter Lagrangian $\Lag_M$ with respect to the tetrad, namely
\begin{eqnarray}
h^{-1}\partial_\sigma \left(h S_a^{\ \rho\sigma} \right)-
h_a^{\ \mu} S_b^{\ \nu\rho} T^b_{\ \nu\mu}+
\frac{h_a^{\ \rho}}{4} T  -  \omega^b_{\ a\sigma}S_b^{\ \rho\sigma}
= \kappa \Theta_a^{\ \rho},\label{femix}
\end{eqnarray}
where as usual we have defined the energy-momentum tensor of matter through
\begin{equation}
\Theta_a^{\ \rho}= \frac{1}{h}\frac{\delta \Lag_M}{\delta h^a_{\ \mu}}. \label{Theta}
\end{equation}

Let us now discuss the Lorentz transformation issues. A local Lorentz transformation is 
represented by the Lorentz matrix $\Lambda^a_{\ b}=\Lambda^a_{\ 
b}(x)$ that obeys
\begin{equation}
\eta_{ab}=\eta_{cd}\,\Lambda^c_{\ a}\Lambda^d_{\ b},
\end{equation}
under which the tetrad and spin connection transform simultaneously as
\begin{equation}
h'{}^a_{\ \mu}=\Lambda^a_{\ b}h^b_{\ \mu}, \qquad \textrm{and}\qquad
\omega'{}^a_{\ b\mu}=\Lambda^a_{\ c}\omega^c_{\ d\mu}\Lambda_b^{\ d}+\Lambda^a_{\ c} 
\partial_\mu \Lambda_b^{\ c}, 
\label{lortrans}
\end{equation}
where $\Lambda_a^{\ b}=(\Lambda^{-1})^b_{\ a}$ is the inverse Lorentz matrix. Hence, if we keep both the tetrad and the spin connection in the 
formulation of TEGR, the theory preserves local Lorentz invariance. 

As one can see, the most general spin connection with vanishing curvature 
is the \textit{purely inertial spin connection} \cite{Pereira.book}
\begin{equation}
\omega^a_{\ b\mu}=\Lambda^a_{\ c} \partial_\mu \Lambda_b^{\ c},
\label{telcon}
\end{equation}
which represents only the inertial effects, since it depends only on the choice of the 
frame.
However, since it depends only on the choice of the observer represented by a Lorentz 
matrix, it is not determined uniquely. In particular, various spin connections represent 
different inertial effects for the observer, which however do not affect the 
field equations \cite{Krssak:2015rqa}. This follows from the fact that the teleparallel 
Lagrangian can be written as 
\cite{Krssak:2015lba}
\begin{equation}
\Lag (h^a_{\ \mu},\omega^a_{\ b\mu})=
\Lag (h^a_{\ \mu},0) +  \frac{1}{\kappa}\partial_\mu \left(
h \omega^\mu
\right), \label{rel}
\end{equation}
where  $\omega^\mu=\omega^{ab}_{\ \ \nu}h_a^{\ \nu}h_{b}^{\ \mu}$. Therefore, the spin 
connection enters the teleparallel action only as a surface term. A variation of the 
surface term vanishes, and hence the choice of the spin connection does not effect the 
field equations. This property allows us to solve  the field equations using an 
arbitrary spin connection. In practice, we usually choose the zero spin connection, which 
is trivially of the correct form (\ref{telcon}).

On the other hand, the torsion tensor (\ref{torr}) is a function of both  the tetrad 
and spin connection. As a consequence, torsion can represent the field strength of both 
gravity and inertia \cite{Krssak:2015lba}. However, there exists a preferred choice of 
the 
spin 
connection, which results in torsion that is the field strength of gravity only. This 
spin 
connection can be obtained by considering the reference tetrad defined by setting the 
gravitational constant to zero
\begin{equation}
\tref^{\;a}{}_{\mu}\equiv\left. h^a_{\ \mu}\right|_{G\rightarrow 0}, \label{reftet}
\end{equation}
and require torsion to vanish there, namely $T^a_{\ 
\mu\nu}(\tref^{\;a}{}_{\mu},\omega^a_{\ 
b\mu})\equiv 0$.
 We find that this defines  the spin connection as 
\begin{equation}
\omega^a_{\ b\mu} = \ombol^a_{\ b\mu}(\tref).
\label{leveq}
\end{equation} 
If the spin connection is chosen in this way the torsion tensor (\ref{torr}) is indeed 
the field strength of gravity only. Consequently, the action (\ref{tegraction}) 
represents the gravitational effects only, without the spurious inertial contributions 
(see \cite{Krssak:2015rqa,Krssak:2015lba} for further details).

We close this section by mentioning the crucial fact that in TEGR the spin connection 
does not effect the field equations, which allows us to use an arbitrary spin connection 
to solve the field equations first, and then calculate the appropriate spin connection 
from the solution of the field equations. This feature lies behind the success of the 
pure tetrad formulation of teleparallel gravity, where the spin connection is considered 
to vanish identically. The solution of the field equations will represent, in addition to 
gravity, the inertial effects too, but these do not affect the metric tensor. Therefore, 
as far as we are  strictly interested in the metric tensor only, the role of the spin 
connection can be neglected. However, this is not the case when one modifies 
the teleparallel Lagrangian and abandons its linearity in the torsion scalar, and indeed 
in this case the covariant formulation is necessary. 

Having developed the techniques for 
covariantizing the theory, in the following section we construct the covariant and fully 
consistent formulation of $f(T)$ gravity, which is the main goal of this work.


\section{Covariant $f(T)$ gravity }
\label{ftsec}

As was mentioned above, in the usual formulation of $f(T)$ gravity one generalizes the 
``pure tetrad teleparallel gravity'' 
\cite{Ferraro:2006jd,Bengochea:2008gz,Ferraro:2008ey,Linder:2010py}, and thus 
the violation of local Lorentz invariance is inherited. However, the crucial novel 
feature is that although in pure tetrad TEGR this violation does not affect the field 
equations, which are still invariant under local Lorentz transformations (which allows us 
to use an arbitrary, for instance zero, spin connection to solve the field equations), in 
the case of ``pure tetrad $f(T)$ gravity'' this is not true anymore and the solutions of 
the field equations do depend on the frame choice. Hence, in the usual formulation of 
$f(T)$ gravity both the action and the field equations are {\it{not}} invariant
under local Lorentz transformations \cite{Li:2010cg}.

In order to clearly see the above problem we recall that according to (\ref{rel}) we can 
write the torsion scalar as
\begin{equation}
T (h^a_{\ \mu},\omega^a_{\ b\mu})=
T (h^a_{\ \mu},0) +  \frac{4}{h}\partial_\mu \left(
h \omega^\mu
\right). \label{holt}
\end{equation} 
Hence, as long as the Lagrangian remains a function linear in $T$, then the spin 
connection appears in the action only as a surface term, and its presence does not affect 
the field equations. However, if the function $f(T)$ is anything different than a linear 
function then the total divergence in (\ref{holt}) is not a total divergence in the whole 
$f(T)$ Lagrangian (\ref{ftaction}). In such a case the variation of the second term will 
be in general non-vanishing, and terms proportional to the spin connection will enter the 
field equations. This implies that the solution of the field equations will depend on the 
choice of the spin connection, which can be understood as that the field equations  are 
potentially the field equations for both gravity and inertia. 

In order to elaborate this problem more, let us recall that the spin connection 
represents just a choice of the observer, i.e. the  inertial effects associated with the 
observer. On the other hand, the  solution of the field equations determines the metric 
tensor, which describes the spacetime geometry. Therefore, we deduce that we are running 
into a severe problem, since the metric tensor of spacetime depends on the inertial 
effects associated with the observer, i.e. the geometry itself becomes frame-dependent.
 
The cause of the problem is that in the usual, pure-tetrad, formulation of $f(T)$ gravity 
the action was constructed using only the first term on the right hand side of 
(\ref{holt}) and the surface term was neglected. However, in the $f(T)$ case, as we have 
just discussed, this neglected term is crucial and the solution of the field equations 
depends on it. Thus, the pure-tetrad $f(T)$ theory leads to physically sensible results 
only in the case where the total divergence in (\ref{holt}) is actually zero, which is 
the case only for the proper tetrad in which the spin connection vanishes. That is why 
people were forced to discuss and construct ``good'' and ``bad'' tetrads 
\cite{Ferraro:2011us,Ferraro:2011ks,Tamanini:2012hg}.

In this section we resolve the above severe problem, by re-formulating $f(T)$ gravity 
starting, instead of the pure-tetrad  teleparallel gravity, from the covariant 
teleparallel gravity presented in the previous section. Hence, we can indeed formulate 
the 
fully covariant, consistent, and frame-independent, version of $f(T)$ gravity, which does 
not suffer from the notorious problems of the usual, pure-tetrad, $f(T)$ theory. 

In order to construct the covariant $f(T)$ gravity, we generalize the covariant 
TEGR. In particular, the action of the theory will be 
 \begin{equation}
\Lag_f=\frac{h}{4\kappa}f(T),
\label{ftaction}
\end{equation}
where $f(T)$ is an arbitrary function of the torsion scalar (\ref{tscalar}). 
The field equations are derived through a variation with respect to the tetrad. As shown 
in detail in the Appendix, they are found to be:
\begin{equation}
E_a^{\ \mu}\equiv
h^{-1} f_T \partial_{\nu}\left( h S_a^{\ \mu\nu} \right)
+
f_{TT} S_a^{\ \mu\nu} \partial_{\nu} T
-
f_T T^b_{\ \nu a }S_b^{\ \nu\mu}
 +
f_T \omega^b_{\ a\nu}S_b^{\ \nu\mu}
+
\frac{1}{4}f(T) h_a^{\ \mu}=\kappa \Theta_a^{\ \mu},
\label{ftequation}
\end{equation}
where $f_T$ and $f_{TT}$ denote first and second order derivatives of $f(T)$ with respect to the torsion scalar $T$.

We mention that  the field equations (\ref{ftequation}) coincide with the  
field equations of the usual $f(T)$ gravity in the case  $\omega^b_{\ a\nu}=0$ \cite{Ferraro:2006jd,Bengochea:2008gz,Ferraro:2008ey,Linder:2010py}.

In summary, the theory (\ref{ftaction}) and the corresponding field equations 
(\ref{ftequation}) are indeed the covariant version of the theory we were looking for. 

\section{Solutions}
\label{solutions}

In the previous section we formulated the covariant $f(T)$ gravity, resulting in the 
covariant field equations (\ref{ftequation}). Nevertheless, we now face the problem of 
how to extract solutions. In particular, although in TEGR the spin connection enters the 
action only as a surface term, and thus we can first solve the field equations to 
determine the tetrad and then calculate the spin connection from the solution, this does 
not hold anymore in the $f(T)$ case where the solution of the field equation does depend 
on the choice of the spin connection but in order to solve the equations we need to have 
the spin connection. This feature implies that in principle we face a loop difficulty. 
 
In order to avoid this problem we need a method of determining the spin connection which 
does not rely on the solution of the field equations. We can recall that in the TEGR case 
the spin connection (\ref{leveq}) is calculated from the knowledge  of the reference 
tetrad  only. Although in the $f(T)$ case it is not possible to determine the reference 
tetrad from the solution of the field equations, the correct reference tetrad can be 
guessed by making some reasonable assumptions based on the symmetries of the geometry. As 
we will show, this is usually possible to be achieved using the knowledge of the 
coordinate system in which the tetrad is written.

In particular, in practice the starting point for any calculation is the ansatz tetrad 
which is given by the symmetry of the problem being investigated. Assuming the knowledge 
of the reference tetrad corresponding to the ansatz tetrad, we can find the spin 
connection using (\ref{leveq}), and then solve the field equations. If our guess of the 
reference tetrad is right then we will be able to remove the inertial effects from 
the action and hence being able to solve the purely gravitational field equations that 
are not contaminated by the spurious inertial contributions. In the following subsections 
we demonstrate this method for various examples.

\subsection{Minkowski Spacetime}

An  illustrative example of the relevance of the spin connection in $f(T)$ theories is 
the simple Minkowski spacetime. We consider two different tetrads representing the 
Minkowski spacetime, each in different coordinate system. Let us start with a diagonal 
tetrad in the Cartesian coordinate system:
\begin{equation}
h^a_{\ \mu}=\text{diag}\left(1,1,1,1\right).
\end{equation}
It is easy to check that this tetrad is a proper tetrad, i.e. the associated inertial 
spin connection vanishes. This can be seen from the fact that the torsion tensor vanishes 
for this tetrad, namely
\begin{equation}
T^a_{\ \mu\nu}(h^a_{\ \mu},0)=0.
\end{equation}
The  field equations in both TEGR and $f(T)$ cases are then trivially satisfied. This is 
an 
expected result, since the field equations should be equations for gravity, which is 
absent in 
Minkowski spacetime.

On the other hand, if we consider a Minkowski diagonal tetrad in the spherical coordinate 
system
\begin{equation}
h^a_{\ \mu}=\text{diag}\left(1,1,r,r\sin\theta\right), \label{minkrad}
\end{equation}
we find that this tetrad is not a proper tetrad, since for zero spin 
connection the corresponding torsion tensor for this tetrad is non-vanishing, namely
\begin{equation}
T^a_{\ \mu\nu}(h^a_{\ \mu},0) \neq 0. \label{tornz}
\end{equation} 
One can clearly see the crucial difference between the  TEGR and $f(T)$ cases. In 
particular, in the TEGR case, despite the fact that (\ref{tornz}) represents the field 
strength of the inertial effects, and consequently the associated Lagrangian is 
non-vanishing
\begin{equation}
\Lag(h^a_{\ \mu},0)=\frac{1}{\kappa}\sin\theta,\label{lagnz}
\end{equation}
the field equations are still satisfied. This is because the holographic relation 
(\ref{rel}) allows the Lagrangian (\ref{lagnz}) to be written as a surface term and 
hence the field equations to be trivially satisfied. On the contrary, this does not 
happen in the $f(T)$ case.

Following the essence of the covariant formulation described in the previous sections, 
the above problem can be solved by calculating an appropriate spin connection, which 
along this tetrad will remove these spurious inertial contributions from the tetrad.
In particular, for the tetrad (\ref{minkrad}), the non-vanishing components of 
the 
inertial spin connection are
\begin{equation}
\omega^{\hat{1}}_{\ \hat{2}\theta}=-1, \quad 
\omega^{\hat{1}}_{\ \hat{3}\phi}=-\sin\theta, \quad
\omega^{\hat{2}}_{\ \hat{3}\phi}=-\cos\theta. \label{spschw}
\end{equation}
If we use this spin connection, the torsion will vanish identically
\begin{equation}
T^a_{\ \mu\nu}(h^a_{\ \mu},\omega^a_{\ b\mu}) = 0,
\end{equation}
and hence the field equations are trivially satisfied in both TEGR and $f(T)$ cases.

\subsection{FRW Universe}

Let us now proceed to the physically more interesting case of a 
Friedmann-Robertson-Walker (FRW) geometry, where one obtains non-trivial gravitational 
field equations.

We start with the  Cartesian coordinate system and the diagonal tetrad that represent the 
FRW metric:
\begin{equation}
h^a_{\ \mu}=\text{diag} (1,a(t),a(t),a(t)), \label{tetradfrwc}
\end{equation}
with $a(t)$ the scale factor (for simplicity we restrict to the flat case, however the 
non-flat case can be studied straightforwardly). The above tetrad leads to the
torsion scalar\footnote{We remind that the superpotential (\ref{superp}) is 
defined without the factor $1/2$ usually used in $f(T)$ gravity literature, and 
therefore the torsion scalar here differs from the usual value by a factor of 2.}
$T=-12 H^2$,
and hence the field equations (\ref{ftequation}) give rise to the Friedmann equations
\begin{eqnarray}
\kappa \rho_M &=& 6 H^2 f_T +\frac{1}{4}f, \label{fried1}\\
\kappa(p_M+\rho_M) &=&2 \dot{H}(24 f_{TT}H^2-f_T),\label{fried2}
\end{eqnarray}
where the subscript ``$T$'' marks the derivative with respect to $T$, $H=\dot{a}/a$ is 
the Hubble parameter, and $\rho_M$ and $p_M$ are the energy density and pressure of the 
matter fluid respectively. These are the correct $f(T)$ modified Friedmann equations 
capable of explaining the accelerated expansion of the Universe, as 
shown in 
\cite{Chen:2010va,Wu:2010mn,Bengochea001,Dent:2011zz,Zheng:2010am,Bamba:2010wb,Cai:2011tc,
Sharif001,Li:2011rn,Capozziello:2011hj,Daouda:2011rt,Wu:2011kh,Wei:2011aa,
Atazadeh:2011aa,Farajollahi:2011af,Karami:2012fu,Cardone:2012xq,Jamil:2012ti,Ong:2013qja,
Amoros:2013nxa,Nesseris:2013jea,Bamba:2013ooa,Kofinas:2014owa,Harko:2014sja,Haro:2014wha,
Geng:2014nfa, Hanafy:2014ica,Darabi:2014dla,Capozziello:2015rda,Nashed:2015pda}. 

On the other hand, in the spherical coordinate system a natural choice for the FRW tetrad 
is the diagonal one
\begin{equation}
h^a_{\ \mu}=\text{diag} \left(1,(1-kr^2)^{-\frac{1}{2}} a, r a,  r a \sin\theta 
\right),\label{tetradfrwl}
\end{equation}
which, similarly to the Minkowski case, is not a proper tetrad. It is easy to check 
that the field equations for this tetrad and vanishing spin connection are satisfied only 
in the TEGR case. 

Similarly to the Minkowski case, this can be easily solved by finding the purely inertial 
spin connection 
corresponding to (\ref{tetradfrwl}). We start with defining the reference tetrad by 
$a(t)=1$, and using (\ref{leveq}) we calculate the non-vanishing components of the spin 
connection as
\begin{eqnarray}
\omega^{\hat{1}}_{\ \hat{2}\theta}=-(1-kr^2)^{\frac{1}{2}},\qquad \omega^{\hat{1}}_{\ \hat{3}\phi}=-(1-kr^2)^{\frac{1}{2}}\sin\theta, 
\qquad \omega^{\hat{2}}_{\ \hat{3}\phi}=-\cos\theta. \label{spfrwl}
\end{eqnarray}
Using this spin connection it is straightforward to check that the tetrad 
(\ref{tetradfrwl}) leads to the same 
field equations (\ref{fried1})-(\ref{fried2}) as the ones obtained  from the tetrad 
(\ref{tetradfrwc}), however we now have the advantage of general covariance. 
This verifies that if we take the role of inertial spin connection into consideration,
then indeed both tetrads are equally good in $f(T)$ gravity, and the theory is not 
frame-dependent anymore.
 

\subsection{Spherically symmetric geometry}

The spherically symmetric spacetime is an important issue that must be addressed properly 
in any theory of gravity. In the framework of $f(T)$ theories it has attracted  lot of 
attention lately 
\cite{Ferraro:2011us,Ferraro:2011ks,Tamanini:2012hg,Wang:2011xf,HamaniDaouda:2011iy,
Nashed:2012ms,Atazadeh:2012am,Nashed:2014sea, 
Aftergood:2014wla,Bhadra:2014jea,delaCruz-Dombriz:2014zaa,Abbas:2015xia,Junior:2015dga, 
Das:2015gwa,Zubair:2015cpa,Ruggiero:2015oka,Nashed:2015pga,Boehmer:2011gw,Gonzalez:2011dr,
Iorio:2012cm,Capozziello:2012zj,Kofinas:2015hla,Kofinas:2015zaa}. In most of these works 
it was argued that only very specific forms of the 
tetrad, with off-diagonal components, can lead to the physical outcome 
\cite{Ferraro:2011us,Ferraro:2011ks,Tamanini:2012hg,Wang:2011xf,HamaniDaouda:2011iy,
Nashed:2012ms,Atazadeh:2012am,Nashed:2014sea, 
Aftergood:2014wla,Bhadra:2014jea,delaCruz-Dombriz:2014zaa,Abbas:2015xia,Junior:2015dga, 
Das:2015gwa,Zubair:2015cpa,Ruggiero:2015oka,Nashed:2015pga}, 
but the physical motivation for this was not understood. Having constructed the covariant 
formulation of $f(T)$ gravity in the previous sections, we can now show that every tetrad 
corresponding to the desired metric, is equally good and leads to solutions, as far as 
the correct spin connection is used. Additionally, we will also provide the explanation 
for the necessity to use the  complicated off-diagonal tetrad in the previous 
non-covariant formulation of $f(T)$ gravity.

As shown in \cite{Dong:2012en}, the spherically symmetric spacetime is necessarily 
static, and 
hence its metric can be written as
\begin{equation}
ds^2=A(r)^2 dt^2-B(r)^2 dr^2-r^2d\theta^2-r^2\sin^2\theta d\phi^2,
\label{metricspherical}
\end{equation}
where $A(r)$ and $B(r)$ are arbitrary functions of the coordinate $r$. The most natural 
choice of the  tetrad that corresponds to this metric has the simple  diagonal form  
\begin{equation}
h^a_{\ \mu}=\text{diag}\left(A(r),B(r),r,r\sin\theta\right)
\label{schwtet}.
\end{equation}
It is straightforward to check that if we assume the trivial spin connection $\omega^a_{\ 
b\mu}=0$, then the field equation $E_{\hat{2}}^{\ \theta}=-\frac{8 f_{TT} 
}{r^5}\cot\theta=0$  gives us necessarily  the condition
$f_{TT}=0$, which restricts the theory to TEGR. In the literature this feature is 
wrongly interpreted as ``the diagonal tetrad is not a good tetrad for 
spherically-symmetric solutions in $f(T)$ gravity'' 
\cite{Ferraro:2011us,Ferraro:2011ks,Tamanini:2012hg,Wang:2011xf,HamaniDaouda:2011iy,
Nashed:2012ms,Atazadeh:2012am,Nashed:2014sea, 
Aftergood:2014wla,Bhadra:2014jea,delaCruz-Dombriz:2014zaa,Abbas:2015xia,Junior:2015dga, 
Das:2015gwa,Zubair:2015cpa,Ruggiero:2015oka,Nashed:2015pga}.

Let us now show that the above issue is an artifact of the non-covariant formulation of 
$f(T)$ gravity. In particular, using the covariant formulation presented in the previous 
sections we will calculate the appropriate spin connection which will allow to use 
any tetrad giving the metric (\ref{metricspherical}), without restricting the functional 
dependence of the Lagrangian. Similarly to 
the previous examples, due to the fact that the solution of the field equations is  
unknown, we have to start with a guess for the reference tetrad corresponding to 
the tetrad (\ref{schwtet}). It is natural to expect that in the absence of 
gravity the diagonal tetrad should reduce to the tetrad (\ref{minkrad}) representing the 
Minkowski spacetime in spherical coordinates. Therefore, the corresponding 
spin connection is again given by 
\begin{equation}
\omega^{\hat{1}}_{\ \hat{2}\theta}=-1, \quad 
\omega^{\hat{1}}_{\ \hat{3}\phi}=-\sin\theta, \quad
\omega^{\hat{2}}_{\ \hat{3}\phi}=-\cos\theta. \label{spschw1}
\end{equation}
Using this spin connection we can now remove the spurious inertial contributions and 
obtain the gravitational torsion tensor. The torsion scalar constructed 
from it is given by
\begin{equation}
T(h^a_{\ \mu},\omega^a_{\ b\mu})=-\frac{4 (-1+B) \left(A-A B+2 r A'\right)}{r^2 A B^2},
\end{equation} 
where primes denote derivative with respect to the radial coordinate $r$. Thus, using 
the tetrad (\ref{schwtet}) and the non-zero 
spin connection (\ref{spschw1}), the field equations (\ref{ftequation}) become
\begin{eqnarray}
 E_{\hat{0}}^{\ t}&\equiv& 
 \left(4 r^4 A B^5\right)^{-1}
\Big[
8 \left(24 f_{TT}-f_{T} r^2\right) B^3
+8 \left(f_{T} r^2-8 f_{TT}\right) B^4+f 
r^4 B^5
\nonumber\\
&& 
+
64 f_{TT}\left( r B'- 2 rB B'+B\right)\Big]
   +
 2\left(r^4 A B^3\right)^{-1}\Big[ \left(8 f_{TT} r+f_{T} r^3\right) B'-24 
f_{TT}\Big]
\nonumber
\\ &&
\left(r^3 A^3 B^5 \right)^{-1}
\Big\{16 f_{TT} r (B-1)^2 B A'^2
+
2 A (B-1) \left[f_{T} r^2 B^3 
A' 
+8 f_{TT} B^2 \left(A'-r A''\right)\right.\nonumber\\
&&    
\left. -16 f_{TT} r A' B'
+8 f_{TT} B 
\left(A' \left(r B'-1\right)+r A''\right)\right]\Big\}
\!=\!0, \label{eq00}
\\
E_{\hat{1}}^{\ r}&\equiv& 
\frac{A \left(8 f_T B-8 f_T+f r^2 B^2\right)+8 f_T r (B-2) 
A'}{4 
r^2 A B^3}
=0,\label{eq11} 
\\
E_{\hat{2}}^{\ \theta}&\equiv&
\left(4 r^5 A^3 
B^5\right)^{-1}
\Big\{
32 f_{TT} r^2 A A' \left\{B^3 A'+2 r A' B'
+B^2 \left(r A''-3 A'\right)\right.\nonumber\\
&&
\left.
-B \left[A' 
\left(r B'-2\right)+r A''\right]\right\}
-32 f_{TT} r^3 (B-1) B A'^3\Big\}
\nonumber
\\
& &
+ \left(4 r^5 B^5\right)^{-1}
\Big[\left(96 f_{TT}-4 f_T r^2\right) B^3
+8 \left(f_T r^2-4 f_{TT}\right) B^4+\left(f 
r^4-4 f_T 
r^2\right) B^5\nonumber\\
&&\ \ \ \ \ 
+32 f_{TT} r B'
+32 f_{TT} B \left(1-2 r B'\right)\Big]+
\frac{ 
 \left(8 f_{TT} r+f_T r^3\right) B'-24 f_{TT}}{r^5 B^3}
 \nonumber 
\\&&
+
 \left(r^4 A B^5\right)^{-1}
 \Big\{ 2 f_T r^2 B^4 A'+24 f_{TT} r A' B'
 -8 f_{TT} B \left[A' 
\left(-2+4 r B'\right)+r A''\right]\Big\}
 \nonumber
\\
& &
+\left(r^4 A B^3\right)^{-1}
\Big\{
16 f_{TT} r A''-  32 f_{TT}A'^2\
+\left(8 f_{TT} r+f_T r^3\right) B'A'
\nonumber\\
&&\ \ \ \ \ 
-
1 B \left[\left(3 f_T r^2-16 f_{TT}\right) A'+r \left(8 f_{TT}+f_T r^2\right) 
A''\right]\Big\}
=0, \label{eq22}
\\
 E_{\hat{3}}^{\ \phi}&\equiv& 
 \frac{1}{\sin\theta}E_{\hat{2}}^{\ \theta}=0. \label{eq33}
\end{eqnarray}
 
As a brief inspection shows, the field equations (\ref{eq00})-(\ref{eq33}) do not 
restrict the form of the $f(T)$-function. Hence, due to the covariant formulation used 
above, the field equations for every $f(T)$-form can be satisfied by all tetrads related 
through Lorentz transformation and corresponding to the spherically-symmetric metric 
(\ref{metricspherical}), and not only by specifically constructed ones.

We stress that the above field equations  (\ref{eq00})-(\ref{eq33}), generated from the 
diagonal tetrad (\ref{schwtet}) and the non-zero 
spin connection (\ref{spschw1}), coincide with those obtained in the usual, 
non-covariant, formulation of $f(T)$ gravity, for zero spin connection but for the 
specific and peculiar non-diagonal tetrad  
\cite{Ferraro:2011us,Ferraro:2011ks,Tamanini:2012hg}
\begin{equation}
\tilde{h}^a_{\ \mu}=
\left( \begin{array}{cccc}
A & 0 & 0 & 0\\ 
0 & B \cos\phi\sin\theta & r\cos\phi\cos\theta & -r \sin\phi\sin\theta \\
0 & -B\cos\theta & r \sin\theta & 0 \\
0 & B\sin\phi\sin\theta & r\sin\phi\cos\theta & r\cos\phi\sin\theta
\end{array}
\right).
\label{offd}
\end{equation} 
This coincidence of the field equations can be easily explained. The off-diagonal tetrad 
(\ref{offd}) is related to the diagonal tetrad (\ref{schwtet}) by a local 
Lorentz transformation of the form
\begin{equation}
\tilde{h}^a_{\ \mu}=\Lambda^a_{\ b} h^b_{\ \mu},
\end{equation}
where the Lorentz matrix is given explicitly by
\begin{equation}
\Lambda^a_{\ b}=
\left( \begin{array}{cccc}
1 & 0 & 0 & 0\\ 
0 & \cos\phi\sin\theta & \cos\phi\cos\theta & -\sin\phi \\
0 & -\cos\theta & \sin\theta & 0 \\
0 & \sin\phi\sin\theta & \sin\phi\cos\theta & \cos\phi
\end{array}
\right).
\label{specLorentz}
\end{equation}
We should now recall that a local Lorentz transformation simultaneously 
transforms both the tetrad and spin connection through  (\ref{lortrans}), and thus the 
spin connection (\ref{spschw1}) gets transformed as well. Interestingly enough, the 
transformed spin connection through (\ref{specLorentz}) is identically zero, namely
\begin{equation}
\tilde{\omega}^a_{\ b\mu}=0.
\end{equation} 
Hence, we can see that the off-diagonal tetrad (\ref{offd}) is a proper tetrad, i.e. 
a tetrad in which the inertial spin connection vanishes, and that is why the obtained 
field equations coincide with the ones of the covariant formulation. In other words, in 
the usual, non-covariant, formulation of $f(T)$ gravity, one considers specific 
peculiar non-diagonal tetrads, and thus making the theory frame-dependent, as a naive way 
to be consistent with a vanishing spin connection. However, as we show, the correct and 
general way to acquire consistency is to use the covariant formulation of $f(T)$ gravity, 
in which case frame-dependence is absent. In particular, one is allowed to use any 
form of the tetrad provided that he calculates the corresponding spin connection. The 
off-diagonal tetrad (\ref{offd}) has no privileged position anymore; it is just a 
specific tetrad in which the corresponding spin connection happens to be zero.

\section{Conclusions}
\label{conclusions}

Taking the serious decision to modify gravity, one still faces the question of which 
formulation of gravity to modify. The usual approach is to start from the curvature-based 
formulation, i.e. general relativity, and modify its action. However, one could start 
from the torsion-based formulation of gravity, i.e from the teleparallel equivalent of 
general relativity (TEGR). The crucial issue is that although TEGR coincides with general 
relativity at the level of equations, their modifications correspond to different 
gravitational theories.

$f(T)$ gravity is the simplest modification of TEGR, as $f(R)$ is the simplest 
modification of GR. However, it is well known that it does not satisfy local Lorentz 
invariance \cite{Li:2010cg}. The reason for that is the following: in the usual 
formulation of $f(T)$ gravity 
\cite{Ferraro:2006jd,Bengochea:2008gz,Ferraro:2008ey,Linder:2010py} one starts from the 
``pure tetrad teleparallel gravity'', i.e he assumes that the spin connection vanishes 
identically, and hence, although the theory becomes simpler, the torsion tensor is 
effectively replaced by the coefficients of anholonomy, that are not tensors under 
local Lorentz transformations. Although in the un-modified gravity, i.e in TEGR, the 
resulting violation of local Lorentz symmetry is often neglected, since it does not 
affect 
the field equations, this becomes a manifest and severe problem in $f(T)$ gravity. 

In this work we solved the above problem, by constructed the consistent, covariant, 
formulation of $f(T)$ gravity. In particular, starting, instead of the pure-tetrad  
teleparallel gravity, from the covariant teleparallel gravity, we were able to 
re-formulate $f(T)$ gravity in a frame-independent way,  which does not suffer from the 
notorious problems of the usual, pure-tetrad, $f(T)$ theory. In such a theory one uses 
both the tetrad and the spin connection, in a way that for every tetrad choice a 
suitably constructed connection makes the whole theory covariant. 

Covariant $f(T)$ gravity is a little bit more involved than the usual, non-covariant one, 
due to the necessity of finding the appropriate spin connection to the tetrad. While in 
covariant TEGR the spin connection enters the action only as a surface term, and thus we 
can first solve the field equations to determine the tetrad and then calculate the spin 
connection from the solution, this does not hold anymore in the $f(T)$ case where 
the solution of the field equation does depend on the choice of the spin connection which 
naively leads to a loop difficulty. In order to avoid this problem we need a method of 
determining the spin connection which does not rely on the solution of the field 
equations. In particular, the correct reference tetrad can be guessed by making some 
reasonable assumptions based on the symmetries of the geometry. Assuming the knowledge 
of the reference tetrad corresponding to the ansatz tetrad, we can find the spin 
connection first, and then solve the field equations. 

As examples we presented the method of extracting solutions in covariant $f(T)$ gravity 
in 
the most physically relevant cases such as the Minkowski, the 
FRW and the spherically-symmetric geometries. As we showed, the field equations for every 
$f(T)$-form can be satisfied by all tetrads related through Lorentz transformation, and 
not only by specifically constructed ones. Hence, there are not ``good'' and ``bad'' 
tetrads in $f(T)$ gravity, there is no-frame dependence, as long as one abandons the 
strong imposition of zero spin connection (the peculiar non-diagonal, ``good'' tetrads 
were just the naive way to be consistent with a vanishing spin connection). 

Covariant $f(T)$ gravity is the correct and consistent way to modify gravity starting 
from its torsion-based formulation. There is an additional difficulty to find the spin 
connection, however we have developed a consistent method to solve this issue. In 
summary, we are able to present a consistent torsional alternative to curvature 
modifications such as $f(R)$ gravity. Hence, covariant 
$f(T)$ gravity and its applications to cosmology and spherically-symmetric geometries,  
must be investigated in detail.

\section*{Acknowledgements}
We would like to thank to Jos\'e G. Pereira for fruitful discussions. This work was 
supported by FAPESP.

\appendix

\section{Derivation of the Field Equations} 

In this Appendix we derive the field equations for the Lagrangian (\ref{ftaction}), 
keeping both the tetrad and the spin connection in the definition of the torsion tensor 
(\ref{torr}). The left hand side of the field equations is given as the Euler-Lagrange 
expression for the  Lagrangian (\ref{ftaction}), namely
\begin{equation}
	E_a^{\ \mu}\equiv \frac{\partial \Lag_{f}}{\partial h^a_{\ \mu}}-
\partial_{\nu}\frac{\partial \Lag_{f}}{\partial (\partial_\nu h^a_{\ \mu})}.
\end{equation}
Up to an overall $1/(4\kappa)$ factor, the first term is given by
\begin{equation}
\frac{\partial h f(T)}{\partial h^a_{\ \mu}}=
h f_T\frac{\partial T }{\partial h^a_{\ \mu}}+
f(T) h h_a^{\ \mu},
\end{equation}
while the second term reads
\begin{eqnarray}
\!\!\!\!\!\!\!
\partial_{\nu}\frac{\partial h f(T)}{\partial (\partial_\nu h^a_{\ \mu})}
=
\partial_{\nu}\left[h f_T \frac{\partial T}{\partial  (\partial_\nu h^a_{\ \mu})} 
\right]
=
f_T \partial_{\nu}\left[ h \frac{\partial T}{\partial  (\partial_\nu h^a_{\ \mu})} 
\right]
+
h(\partial_{\nu} T)  f_{TT} \frac{\partial T}{\partial  (\partial_\nu h^a_{\ \mu})} 
.
\end{eqnarray}
The derivatives of the torsion scalar are well-known in the ordinary TEGR case (for 
instance see the Appendix C in \cite{Pereira.book}), namely
\begin{eqnarray}
\frac{\partial T}{\partial  (\partial_\nu h^a_{\ \mu})}&=& - 4 S_a^{\ \mu\nu}
\\
\frac{\partial T}{\partial  ( h^a_{\ \mu})}&=&
-4 T^b_{\ \nu a }S_b^{\ \nu\mu}+ 4 \omega^b_{\ a\nu}S_b^{\ \nu\mu}.
\end{eqnarray}
Assembling the above we finally obtain
\begin{eqnarray}
&&\!\!\!\!\!\!\!\!\!\!\!\!\!\!\!\!\!
E_a^{\ \mu}=
h^{-1} f_T \partial_{\nu}\left( h S_a^{\ \mu\nu} \right)
+
f_{TT} S_a^{\ \mu\nu} \partial_{\nu} T
-
f_T T^b_{\ \nu a }S_b^{\ \nu\mu}
 +
f_T \omega^b_{\ a\nu}S_b^{\ \nu\mu}
+
\frac{1}{4}f(T) h_a^{\ \mu}.
\label{ftequation00}
\end{eqnarray}
Hence, the field equations write as
\begin{eqnarray}
E_a^{\ \mu}=\kappa \Theta_a^{\ \mu},
\label{ftequation01}
\end{eqnarray}
where as usual we have defined the energy-momentum tensor of matter through
\begin{equation}
\Theta_a^{\ \rho}= \frac{1}{h}\frac{\delta \Lag_M}{\delta h^a_{\ \mu}}. \label{Theta00}
\end{equation}


\begin{thebibliography}{99}
 

 
 
\bibitem{Nojiri:2006ri}
  S.~Nojiri and S.~D.~Odintsov,
  eConf {\bf C0602061}, 06 (2006), Int.\ J.\ Geom.\ Meth.\ Mod.\ Phys.\ 
{\bf 4}, 115 (2007).


\bibitem{Capozziello:2011et}
  S.~Capozziello and M.~De Laurentis,
  Phys.\ Rept.\  {\bf 509}, 167 (2011).
 
 
\bibitem{Copeland:2006wr}
  E.~J.~Copeland, M.~Sami and S.~Tsujikawa,
  Int.\ J.\ Mod.\ Phys.\  D {\bf 15}, 1753 (2006).

 
 

\bibitem{Cai:2009zp}
  Y.~F.~Cai, E.~N.~Saridakis, M.~R.~Setare and J.~Q.~Xia,
\textit{Quintom Cosmology: Theoretical implications and observations},
  Phys.\ Rept.\  {\bf 493}, 1 (2010).

 
  \bibitem {Stelle:1976gc}
  K.~S.~Stelle,
Phys.\ Rev.\ D \textbf{16}, 953 (1977).
 
\bibitem{Biswas:2011ar} 
  T.~Biswas, E.~Gerwick, T.~Koivisto and A.~Mazumdar,
  Phys.\ Rev.\ Lett.\  {\bf 108}, 031101 (2012).

  

\bibitem{DeFelice:2010aj}
A.~De Felice and S.~Tsujikawa,
Living Rev.\ Rel.\  {\bf 13} (2010) 3.

\bibitem{Nojiri:2010wj} 
  S.~Nojiri and S.~D.~Odintsov,
  Phys.\ Rept.\  {\bf 505}, 59 (2011).



\bibitem{ein28}
A. Einstein 1928, Sitz. Preuss. Akad. Wiss. p. 217; ibid p. 224.


\bibitem{Unzicker:2005in}
  A.~Unzicker and T.~Case,
  physics/0503046.

  
\bibitem{Sauer}
T.~Sauer, 
Historia Math. 33 (2006) 399-439.

  
  \bibitem{Moller}
C.~M\o ller,
K. Dan. Vidensk. Selsk. Mat. Fys. Skr. \textbf{1}, 10 (1961). 


\bibitem{Hayashi:1967se}
K.~Hayashi and T.~Nakano,
Prog.\ Theor.\ Phys.\  {\bf 38} (1967) 491.

\bibitem{Cho:1975dh}
Y.~M.~Cho,
Phys.\ Rev.\ D {\bf 14}, 2521 (1976).



\bibitem{Hayashi:1977jd}
K.~Hayashi,
Phys.\ Lett.\ B {\bf 69} (1977) 441.


  

 \bibitem{Hayashi79}
  K. Hayashi and T. Shirafuji,
  Phys. Rev. D \textbf{19}, 3524 (1979);
  Addendum-ibid. \textbf{24}, 3312 (1982).


\bibitem{Maluf:1994ji}
  J.~W.~Maluf,
  J.\ Math.\ Phys.\ \textbf{35} (1994) 335.
  
\bibitem{Arcos:2005ec}
  H.~I.~Arcos and J.~G.~Pereira,
  Int.\ J.\ Mod.\ Phys.\ D \textbf{13}, 2193 (2004).





 \bibitem{Pereira.book}
 R. Aldrovandi, J.G. Pereira,
{\it{Teleparallel Gravity: An Introduction}},
Springer, Dordrecht, 2013.

 
\bibitem{Ferraro:2006jd}
  R.~Ferraro and F.~Fiorini,
  Phys.\ Rev.\ D {\bf 75}, 084031 (2007).
  
\bibitem{Bengochea:2008gz} 
  G.~R.~Bengochea and R.~Ferraro,
  Phys.\ Rev.\ D {\bf 79}, 124019 (2009).
  
\bibitem{Ferraro:2008ey} 
  R.~Ferraro and F.~Fiorini,
  Phys.\ Rev.\ D {\bf 78}, 124019 (2008).
 

\bibitem{Linder:2010py}
  E.~V.~Linder,
  Phys.\ Rev.\ D \textbf{81}, 127301 (2010).
 
  
    
\bibitem{Chen:2010va}
S.~H.~Chen, J.~B.~Dent, S.~Dutta and E.~N.~Saridakis,
Phys.\ Rev.\ D
\textbf{ 83}, 023508 (2011).

\bibitem{Wu:2010mn}
  P.~Wu, H.~W.~Yu,
  Phys.\ Lett.\ \textbf{B693}, 415 (2010).


\bibitem{Bengochea001}
    G.~R.~Bengochea,
  Phys.\ Lett.\  {\bf B695}, 405 (2011).

\bibitem{Dent:2011zz} 
J.~B.~Dent, S.~Dutta, E.~N.~Saridakis,
  JCAP {\bf 1101}, 009 (2011).

\bibitem{Zheng:2010am}
  R.~Zheng and Q.~G.~Huang,
  JCAP {\bf 1103}, 002 (2011).

\bibitem{Bamba:2010wb}
  K.~Bamba, C.~Q.~Geng, C.~C.~Lee and L.~W.~Luo,
  JCAP {\bf 1101}, 021 (2011).

\bibitem{Cai:2011tc}
  Y.~-F.~Cai, S.~-H.~Chen, J.~B.~Dent, S.~Dutta, E.~N.~Saridakis,
  Class.\ Quant.\ Grav.\  {\bf 28}, 215011 (2011).

\bibitem{Sharif001}
  M.~Sharif, S.~Rani,
  Mod.\ Phys.\ Lett.\  {\bf A26}, 1657 (2011).

\bibitem{Li:2011rn}
  M.~Li, R.~X.~Miao and Y.~G.~Miao,
  JHEP {\bf 1107}, 108 (2011).

\bibitem{Capozziello:2011hj}
  S.~Capozziello, V.~F.~Cardone, H.~Farajollahi and A.~Ravanpak,
  Phys.\ Rev.\  D {\bf 84}, 043527 (2011).

\bibitem{Daouda:2011rt}
 M.~H. Daouda, M.~E.~Rodrigues and M.~J.~S.~Houndjo,
  Eur.\ Phys.\ J.\ C {\bf 72}, 1890 (2012).
  
\bibitem{Wu:2011kh}
  Y.~P.~Wu and C.~Q.~Geng,
     Phys.\ Rev.\ D {\bf 86}, 104058 (2012).

\bibitem{Wei:2011aa}
  H.~Wei, X.~J.~Guo and L.~F.~Wang,
  Phys.\ Lett.\  B {\bf 707}, 298 (2012).

\bibitem{Atazadeh:2011aa}
  K.~Atazadeh and F.~Darabi,
 Eur.Phys.J. C72 (2012) 2016.

\bibitem{Farajollahi:2011af}
  H.~Farajollahi, A.~Ravanpak and P.~Wu,
  Astrophys.\ Space Sci.\  {\bf 338}, 23 (2012).

\bibitem{Karami:2012fu}
  K.~Karami and A.~Abdolmaleki,
 JCAP 1204 (2012) 007.

\bibitem{Cardone:2012xq} 
  V.~F.~Cardone, N.~Radicella and S.~Camera,
  Phys.\ Rev.\ D {\bf 85}, 124007 (2012).

 \bibitem{Jamil:2012ti}
  M.~Jamil, D.~Momeni and R.~Myrzakulov,
  Eur.\ Phys.\ J.\ C {\bf 72}, 2267 (2012).
 
\bibitem{Ong:2013qja}
  Y.~C.~Ong, K.~Izumi, J.~M.~Nester and P.~Chen,
  Phys.\ Rev.\ D {\bf 88} (2013) 2,  024019.

 \bibitem{Amoros:2013nxa}
  J.~Amoros, J.~de Haro and S.~D.~Odintsov,
     Phys.\ Rev.\ D {\bf 87}, 104037 (2013).

\bibitem{Nesseris:2013jea} 
  S.~Nesseris, S.~Basilakos, E.~N.~Saridakis and L.~Perivolaropoulos,
  Phys.\ Rev.\ D {\bf 88}, 103010 (2013).

\bibitem{Bamba:2013ooa} 
  K.~Bamba, S.~Capozziello, M.~De Laurentis, S.~'i.~Nojiri and D.~S\'aez-G\'omez,
  Phys.\ Lett.\ B {\bf 727}, 194 (2013).

\bibitem{Kofinas:2014owa} 
  G.~Kofinas and E.~N.~Saridakis,
  Phys.\ Rev.\ D {\bf 90}, 084044 (2014).

  
\bibitem{Harko:2014sja} 
  T.~Harko, F.~S.~N.~Lobo, G.~Otalora and E.~N.~Saridakis,
  Phys.\ Rev.\ D {\bf 89}, 124036 (2014).

\bibitem{Haro:2014wha} 
  J.~Haro and J.~Amoros,
 JCAP {\bf 1412} (2014) 12,  031.

\bibitem{Geng:2014nfa} 
  C.~Q.~Geng, C.~Lai, L.~W.~Luo and H.~H.~Tseng,
  Phys.\ Lett.\ B {\bf 737}, 248 (2014).

  
\bibitem{Hanafy:2014ica} 
  W.~El Hanafy and G.~G.~L.~Nashed,
  Eur.\ Phys.\ J.\ C {\bf 75}, 279 (2015).

\bibitem{Darabi:2014dla} 
  F.~Darabi, M.~Mousavi and K.~Atazadeh,
  Phys.\ Rev.\ D {\bf 91}, 084023 (2015).

\bibitem{Capozziello:2015rda} 
  S.~Capozziello, O.~Luongo and E.~N.~Saridakis,
 Phys.\ Rev.\ D {\bf 91} (2015) 12,  124037.

  
\bibitem{Nashed:2015pda} 
  G.~L.~Nashed,
   Gen.\ Rel.\ Grav.\  {\bf 47} (2015) 7,  75.

   


\bibitem{Li:2010cg}
B.~Li, T.~P.~Sotiriou and J.~D.~Barrow,
Phys.\ Rev.\ D {\bf 83} (2011) 064035.

\bibitem{Ferraro:2014owa}
  R.~Ferraro and F.~Fiorini,
  Phys.\ Rev.\ D {\bf 91} (2015) 6,  064019.

  
\bibitem{Bahamonde:2015zma}
  S.~Bahamonde, C.~G.~Boehmer and M.~Wright,
  Phys.\ Rev.\ D {\bf 92} (2015) 10,  104042.



\bibitem{Junior:2015kia}
  E.~L.~B.~Junior and M.~E.~Rodrigues,
  arXiv:1509.03267 [gr-qc].



\bibitem{Obukhov:2006sk}
Y.~N.~Obukhov and G.~F.~Rubilar,
Phys.\ Rev.\ D {\bf 73} (2006) 124017.




\bibitem{Maluf:2013gaa}
J.~W.~Maluf,
Ann. Phys.\ (Berlin)  {\bf 525} (2013) 339.



 






\bibitem{Obukhov:2002tm}
Y.~N.~Obukhov and J.~G.~Pereira,
Phys.\ Rev.\ D {\bf 67} (2003) 044016.


\bibitem{Hehl:1994ue}
  F.~W.~Hehl, J.~D.~McCrea, E.~W.~Mielke and Y.~Ne'eman,
  Phys.\ Rept.\  {\bf 258} (1995) 1.

\bibitem{Lucas:2009nq}
T.~G.~Lucas, Y.~N.~Obukhov and J.~G.~Pereira,
Phys.\ Rev.\ D {\bf 80} (2009) 064043.
    
     

\bibitem{Krssak:2015rqa}
  M.~Kr\v{s}\v{s}\'ak and J.~G.~Pereira,
Eur.\ Phys.\ J.\ C {\bf 75} (2015) 519.

  
\bibitem{Krssak:2015lba}
  M.~Kr\v{s}\v{s}\'ak,
  arXiv:1510.06676 [gr-qc].
 
  
\bibitem{Ferraro:2011us} 
  R.~Ferraro and F.~Fiorini,
  Phys.\ Lett.\ B {\bf 702}, 75 (2011).

  
\bibitem{Ferraro:2011ks} 
  R.~Ferraro and F.~Fiorini,
  Phys.\ Rev.\ D {\bf 84}, 083518 (2011).


\bibitem{Tamanini:2012hg}
  N.~Tamanini and C.~G.~Boehmer,
  Phys.\ Rev.\ D {\bf 86} (2012) 044009.
  
\bibitem{Wang:2011xf}
  T.~Wang,
  Phys.\ Rev.\ D {\bf 84}, 024042 (2011).

\bibitem{HamaniDaouda:2011iy}
  M.~Hamani Daouda, M.~E.~Rodrigues and M.~J.~S.~Houndjo,
  Eur.\ Phys.\ J.\ C {\bf 71}, 1817 (2011).

  
\bibitem{Nashed:2012ms}
  G.~G.~L.~Nashed,
Eur.\ Phys.\ J.\ C {\bf 73} (2013) 4,  2394.

\bibitem{Atazadeh:2012am}
  K.~Atazadeh and M.~Mousavi,
  Eur.\ Phys.\ J.\ C {\bf 73} (2013) 1,  2272.
 
\bibitem{Nashed:2014sea}
  G.~G.~L.~Nashed,
  Europhys.\ Lett.\  {\bf 105}, 10001 (2014).
 
\bibitem{Aftergood:2014wla}
  J.~Aftergood and A.~DeBenedictis,
  Phys.\ Rev.\ D {\bf 90} (2014) 12,  124006.
 
\bibitem{Bhadra:2014jea}
  J.~Bhadra and U.~Debnath,
  Int.\ J.\ Theor.\ Phys.\  {\bf 53}, 645 (2014).

  
\bibitem{delaCruz-Dombriz:2014zaa}
  \'A.~de la Cruz-Dombriz, P.~K.~S.~Dunsby and D.~Saez-Gomez,
  JCAP {\bf 1412}, 048 (2014).
  
\bibitem{Abbas:2015xia}
  G.~Abbas, S.~Qaisar and M.~A.~Meraj,
  Astrophys.\ Space Sci.\  {\bf 357} (2015) 2,  156.
 
\bibitem{Junior:2015dga}
  E.~L.~B.~Junior, M.~E.~Rodrigues and M.~J.~S.~Houndjo,
  JCAP {\bf 1506} (2015) 06,  037.
 
\bibitem{Das:2015gwa}
  A.~Das, F.~Rahaman, B.~K.~Guha and S.~Ray,
  Astrophys.\ Space Sci.\  {\bf 358} (2015) 2,  36.
  
\bibitem{Zubair:2015cpa}
  M.~Zubair and G.~Abbas,
Astrophys.\ Space Sci.\  {\bf 361} (2016) 1,  27.
 
\bibitem{Ruggiero:2015oka}
  M.~L.~Ruggiero and N.~Radicella,
  Phys.\ Rev.\ D {\bf 91}, 104014 (2015).
  

\bibitem{Nashed:2015pga}
  G.~G.~L.~Nashed,
   Eur.\ Phys.\ J.\ Plus {\bf 130} (2015) 7,  124.

  
 
 
\bibitem{Boehmer:2011gw}
  C.~G.~Boehmer, A.~Mussa and N.~Tamanini,
  Class.\ Quant.\ Grav.\  {\bf 28}, 245020 (2011).

\bibitem{Gonzalez:2011dr}
  P.~A.~Gonzalez, E.~N.~Saridakis and Y.~Vasquez,
  JHEP {\bf 1207}, 053 (2012).
 
 
 
\bibitem{Iorio:2012cm} 
  L.~Iorio and E.~N.~Saridakis,
  Mon.\ Not.\ Roy.\ Astron.\ Soc.\  {\bf 427}, 1555 (2012).

 
\bibitem{Capozziello:2012zj}
  S.~Capozziello, P.~A.~Gonzalez, E.~N.~Saridakis and Y.~Vasquez,
  JHEP {\bf 1302}, 039 (2013).
 
 
\bibitem{Kofinas:2015hla}
  G.~Kofinas, E.~Papantonopoulos and E.~N.~Saridakis,
  Phys.\ Rev.\ D {\bf 91}, 104034 (2015).
 
\bibitem{Kofinas:2015zaa}
  G.~Kofinas,
   Phys.\ Rev.\ D {\bf 92} (2015) 8,  084022.
 
  
  
\bibitem{Dong:2012en}
  H.~Dong, Y.~b.~Wang and X.~h.~Meng,
  Eur.\ Phys.\ J.\ C {\bf 72} (2012) 2002.
  
\end{thebibliography}
\end{document}